\documentclass[final,3p]{elsarticle}
\biboptions{numbers,sort&compress}

\usepackage{amssymb}
\usepackage[]{siunitx}
\usepackage{multirow}
\usepackage{wasysym}
\usepackage{xcolor}
\usepackage{xspace}
\usepackage{subcaption}
\usepackage{graphicx}
\usepackage[utf8]{inputenc}
\usepackage{makecell}
\usepackage[page,toc,titletoc,title]{appendix}
\usepackage[subfigure]{tocloft}
\usepackage[section]{placeins} 
\usepackage{lineno}

\usepackage[export]{adjustbox}

\usepackage{siunitx} 
\DeclareSIUnit\sq{\ensuremath{\Box}}                           

\newcommand{\smartpixels}{\emph{smartpixels}\xspace}

\usepackage[rawfloats=true]{floatrow}

\usepackage{xcolor}

\usepackage{hyperref}

\journal{Nucl. Instrum. Meth. A}

\begin{document}

\begin{frontmatter}


\title{Robustness of the \textit{smartpixels} classifier for different simulated sensor geometries and non-ideal detector conditions} 


\author[a]{Danush Shekar\corref{cor1}}\ead{dsheka3@uic.edu}
\author[b]{Ben Weiss}
\author[c]{Morris Swartz}
\author[a]{Corrinne Mills}
\author[b]{Jennet Dickinson}
\author[d]{Lindsey Gray}
\author[e]{David Jiang}
\author[a]{Mohammad Abrar Wadud}
\author[f]{Daniel Abadjiev}
\author[f]{Anthony Badea}
\author[d]{Douglas Berry}
\author[b]{Alec Cauper}
\author[h]{Arghya Ranjan Das}
\author[f]{Karri Folan DiPetrillo}
\author[d,k]{Farah Fahim}
\author[f]{Rachel Kovach Fuentes}
\author[d]{Abhijith Gandrakota}
\author[d,k]{Giuseppe Di Guglielmo}
\author[f]{Eliza Howard}
\author[h]{Shiqi Kuang}
\author[f]{Carissa Kumar}
\author[h]{Mia Liu}
\author[c]{Petar Maksimovic}
\author[l]{Nick Manganelli}
\author[e]{Mark S Neubauer}
\author[f]{Aidan Nicholas}
\author[f]{Emily Pan}
\author[d]{Benjamin Parpillon}
\author[i]{Jannicke Pearkes}
\author[i]{Ricardo Silvestre}
\author[d]{Chinar Syal}
\author[a]{Amit Trivedi}
\author[i]{Keith Ulmer}
\author[a]{Jieun Yoo}
\author[f]{Eric You}

\affiliation[a]{University of Illinois at Chicago, Chicago, IL 60607, USA}
\affiliation[b]{Cornell University, Ithaca, NY 14853, USA}
\affiliation[c]{Johns Hopkins University, Baltimore, MD 21218, USA}
\affiliation[d]{Fermi National Accelerator Laboratory, Batavia, IL 60510, USA}
\affiliation[e]{University of Illinois Urbana-Champaign, Champaign, IL 61801, USA}
\affiliation[f]{The University of Chicago, Chicago, IL 60637, USA}
\affiliation[h]{Purdue University, West Lafayette, IN 47907, USA}
\affiliation[l]{Northeastern University, Boston, MA 02115, USA} 
\affiliation[i]{University of Colorado Boulder, Boulder, CO 80309, USA}
\affiliation[k]{Northwestern University, Evanston, IL 60208, USA}

\begin{abstract}
Pixel tracking detectors at upcoming collider experiments will see unprecedented charged-particle densities.  Real-time data reduction on the detector will enable higher granularity and faster readout, possibly enabling the use of the pixel detector in high-rate online event selection, such as the ATLAS or CMS first-level trigger systems.  This data reduction can be accomplished with a neural network (NN) in the readout chip bonded with the sensor that recognizes and rejects tracks with low transverse momentum (p$_T$) based on the geometrical shape of the charge deposition (``cluster'').  To design viable detectors for deployment, the dependence of the NN as a function of the sensor geometry, external magnetic field, irradiation, and noise must be understood.  In this paper, we present first studies of the efficiency and data reduction for planar pixel sensors exploring these parameters.  For the CMS HL-LHC sensor geometry, we obtain a signal efficiency of (91.9 $\pm$ 0.7)\% and a data reduction of (29.7 $\pm$ 1.0)\%. A smaller sensor pitch in the bending direction improves the p$_T$ discrimination, but a larger pitch can be partially compensated with detector thickness.    Any accumulated radiation damage also changes the cluster shape, reducing the signal efficiency compared to the baseline by approximately 30\textendash{}60\% in absolute terms, but nearly all of the performance can be recovered through retraining of the network and updating the weights. Finally, the impact of noise was investigated, and retraining the network on noise-injected datasets was found to maintain performance within 6\% of the baseline network trained and evaluated on noiseless data.

\end{abstract}




\end{frontmatter}
\tableofcontents


\section{Introduction}\label{sec:introduction}

Next-generation collider detectors will see unprecedented collision rates and particle densities, putting severe pressure on detector readout bandwidth.  The track density is highest near the collision point, where, conventionally, highly granular silicon ``pixel'' tracking detectors are placed, as in the CMS~\cite{CMS_2008} and ATLAS~\cite{ATLAS_2008} experiments at CERN. That increased density motivates ever smaller pitch in the detector design, which further exacerbates the bandwidth needs, even with zero-suppressed readout. The \smartpixels project is designing machine learning (ML) algorithms and application-specific integrated circuits (ASICs) that can be deployed on a pixel detector to perform inference and data reduction on-detector, reducing the bandwidth requirements. As an example, the \smartpixels technology introduced to the innermost sensors in the CMS and ATLAS experiments would help unlock new physics capabilities by facilitating the potential inclusion of inner-tracker data in the Level 1 trigger. Improved tracking performance, particularly in the resolution of particle impact parameters, are expected to yield the largest increases in trigger efficiencies for signatures with heavy quarks, particularly the flagship $hh\to{4b}$ analysis, hadronic top quark decays for measurements and searches, and key $b$ physics measurements, but will likely also be useful in searches for exotic phenomena including long-lived particles.


The \smartpixels project established the feasibility of filtering the tracks of incident particles by transverse momentum though a neural network implementable on an ASIC~\cite{Yoo_2024}.  The filtering algorithm is a classifier that uses the shape of the cluster of adjacent pixels that see a signal from the particle crossing the detector.  An ASIC with this algorithm implemented has been produced and characterized in the laboratory~\cite{BadeaParpillon_2024}.  On the algorithm side, further work explores the possibility of regressing the incident track parameters and their uncertainties using the cluster shape~\cite{Dickinson_2023}.  

A necessary step to the deployment of \smartpixels in a collider detector is understanding how the physical design choices affect the performance of the algorithm(s) implemented.  This paper therefore investigates how sensor design parameters and operating conditions such as pixel pitch, active thickness, Lorentz drift, radiation damage, and noise impact performance, using the classifier network architecture developed in~\cite{Yoo_2024} as a reference benchmark.  

The datasets to test and train the filtering algorithm are defined in Section~\ref{sec:datasets}, and Section~\ref{sec:filter}  reviews the structure and training of the filter algorithm.  Then, we show the performance of the filter algorithm for different pixel pitches in both transverse directions and sensor thickness (Section~\ref{sec:geometry}), for sensors oriented in the magnetic field so that charges have no Lorentz drift (Section~\ref{sec:lorentz}), for sensors that have been irradiated to fluences typical of the High-Luminosity Large Hadron Collider (HL-LHC) (Section~\ref{sec:irradiation}), and for devices with noise from the ASIC based on simulated detector conditions (Section~\ref{sec:noise}).


\section{Simulated datasets}\label{sec:datasets}
The PixelAV~\cite{Swartz_pixelAV} program is used to generate the datasets required for training and validating the filtering network. PixelAV provides a detailed model of silicon pixel detectors, including drift, diffusion, and the Hall effect.  It is also capable of simulating radiation effects such as charge trapping and the signals induced from trapped charges.  Electric and weighting field maps of the sensor designs were simulated using Silvaco TCAD~\cite{silvaco_tcad} and sent as input to PixelAV. The version of PixelAV used to generate the datasets in this paper also provides granular information on the time evolution of signals in the sensor, and the datasets here record 20 ``slices'' in steps of 200 \unit{\pico\second}.  PixelAV calculates the signals induced by incident charged pions in an array of 21$\times$13 pixels based on the trajectory of the incident particles.  For each event, a single charged particle is simulated, with its incident position ($x$, $y$) on the sensor mid-plane uniformly distributed across the central 3$\times$3 pixels of the 21$\times$13 array. We neglect possible edge effects in this work, anticipating the development of a dynamic approach to centering the window of pixels input to the NN on the cluster.  Given the large number of pixels in a modern sensor and ASIC, the effect of the physical edge of the sensor is considered negligible for these studies. The collection of datasets considered for training and validation has been expanded compared to the ones generated for the previous study~\cite{Yoo_2024}.

The baseline performance of the filtering algorithm~\cite{Yoo_2024} was evaluated using simulated datasets from a silicon planar sensor (pitch = 50$\times$12.5 $\unit{\micro\meter}^2$ along $x\times y$ and thickness = 100 $\unit{\micro\meter}$ along $z$) assumed to be in the barrel region of the CMS experiment. The incident particle makes an angle with the sensor array in the $x - z$ and $y - z$ planes denoted by $\alpha$ and $\beta$, respectively. The trajectories of particles emanating from the interaction point at the origin curve due to the 3.8~T external magnetic field along $x$ before encountering the sensor array placed along a cylinder at approximately 30 \unit{\milli\meter} radius from the origin. Due to that curvature, the shape of the cluster depends on the particle’s transverse momentum, which is highly correlated with $\beta$. In particular, the cluster shape along the $y$-direction is sensitive to the incident angle $\beta$ as well as the p$_T$ of the particle. The size of the pixel array by choice is large enough to fully encompass a cluster for these incident positions and angles ($\alpha$, $\beta$).

\begin{figure}[tbh]
    \centering
    \includegraphics[width=0.6\linewidth]{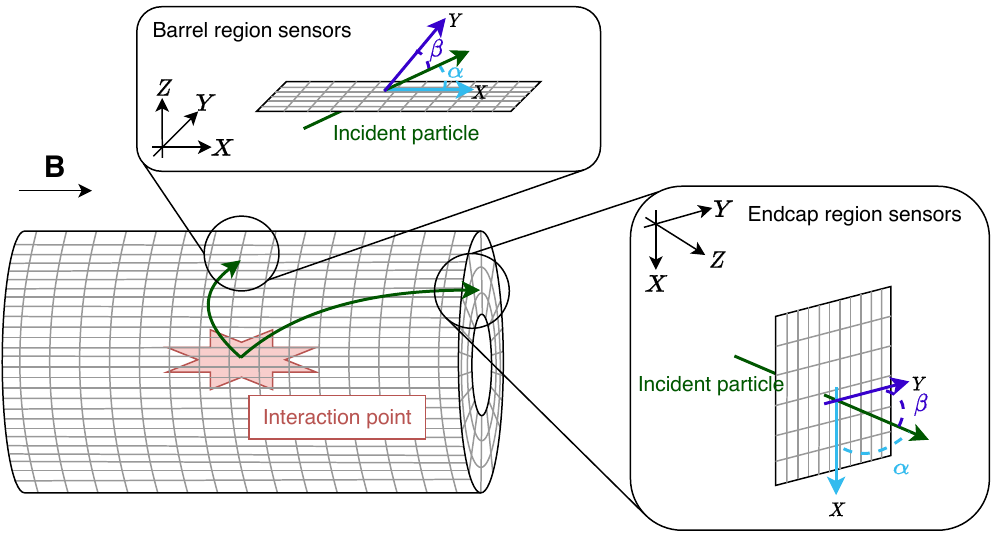}
    \caption{A schematic diagram of two arbitrary charged particles traversing sensors in the barrel and endcap regions, showing the magnetic field direction and local coordinate systems. The convention for the local coordinate system is that the longer and shorter sides of the pixel pitch are along the $x$-direction and $y$-direction, respectively.  The angle between the incident particle's projection in the $x$-$z$ plane and the $x$-axis at the sensor is labeled $\alpha$, and the angle between its projection in the $y$-$z$ plane and the $y$-axis is labeled $\beta$.  The local coordinate system changes with respect to the global coordinate system to follow the local sensor geometry.}
    \label{fig:geometry-definitions}
\end{figure}

There are typically regions in collider experiments, like the endcap pixel sensors in the CMS experiment, where sensors are oriented such that the external magnetic field is parallel to the sensor electric field. In such scenarios, there is no Lorentz drift (deflection of charge carriers by the magnetic field) that otherwise improves impact position reconstruction via interpolation. Thus, we include simulations of sensors in endcap regions. 
While the orientation of the sensors in the endcaps would be different with respect to the barrel region, $\alpha$ and $\beta$ are still defined as the angles made with the sensor length-thickness ($x - z$) and sensor width-thickness ($y - z$) planes, respectively. This definition is maintained across both detector regions by transforming the coordinate systems as illustrated in Figure~\ref{fig:geometry-definitions}. All results presented in this paper are expressed in natural units.

The distribution of induced charges in the pixel array forms the bulk of input information sent to the classification network, and thus, the granularity of the pixel pitch and the sensor thickness influence the network's performance. To understand this impact, five sensor geometries that vary in pitch along $y$ as listed in Table~\ref{tab:sensor-list} were investigated. To offer comparison to a specific detector design, we include in the survey a sensor matching the inner tracker sensor geometry in the CMS Phase-2 upgrade (100$\times$25$\times$150~$\unit{\micro\meter}^3$)~\cite{CMS:2017lum}, along with a 100 $\unit{\micro\meter}$ thick variant of the same.

\begin{table}[h!]
\centering
\resizebox{4in}{!}{%
\begin{tabular}{|c|c|ccc|c|}
\hline
\multirow{2}{*}{Sl. No.} &
  \multirow{2}{*}{\begin{tabular}[c]{@{}c@{}}Sensor \\ name\end{tabular}} &
  \multicolumn{3}{c|}{Sensor geometry} &
  \multirow{2}{*}{\begin{tabular}[c]{@{}c@{}}Bias voltage\\ {[}V{]}\end{tabular}} \\ \cline{3-5}
 &
   &
  \multicolumn{1}{c|}{\begin{tabular}[c]{@{}c@{}}Length\\ {[}$\unit{\micro\meter}${]}\end{tabular}} &
  \multicolumn{1}{c|}{\begin{tabular}[c]{@{}c@{}}Width\\ {[}$\unit{\micro\meter}${]}\end{tabular}} &
  \begin{tabular}[c]{@{}c@{}}Thickness\\ {[}$\unit{\micro\meter}${]}\end{tabular} &
   \\ \hline
1 & S1 & \multicolumn{1}{c|}{50}  & \multicolumn{1}{c|}{10}   & 100 & 100 \\ \hline
2 & S2 & \multicolumn{1}{c|}{50}  & \multicolumn{1}{c|}{12.5} & 100 & 100 \\ \hline
3 & S3 & \multicolumn{1}{c|}{50}  & \multicolumn{1}{c|}{15}   & 100 & 100 \\ \hline
4 & S4 & \multicolumn{1}{c|}{50}  & \multicolumn{1}{c|}{20}   & 100 & 100 \\ \hline
5 & S5 & \multicolumn{1}{c|}{50}  & \multicolumn{1}{c|}{25}   & 100 & 100 \\ \hline
6 & S6 & \multicolumn{1}{c|}{100} & \multicolumn{1}{c|}{25}   & 100 & 100 \\ \hline
7 & S7 & \multicolumn{1}{c|}{100} & \multicolumn{1}{c|}{25}   & 150 & 175 \\ \hline
\end{tabular}%
}
\caption{Specifications of the sensor geometries surveyed.}
\label{tab:sensor-list}
\end{table}

\begin{figure}[h]
    \centering
    \includegraphics[width=0.32\textwidth]{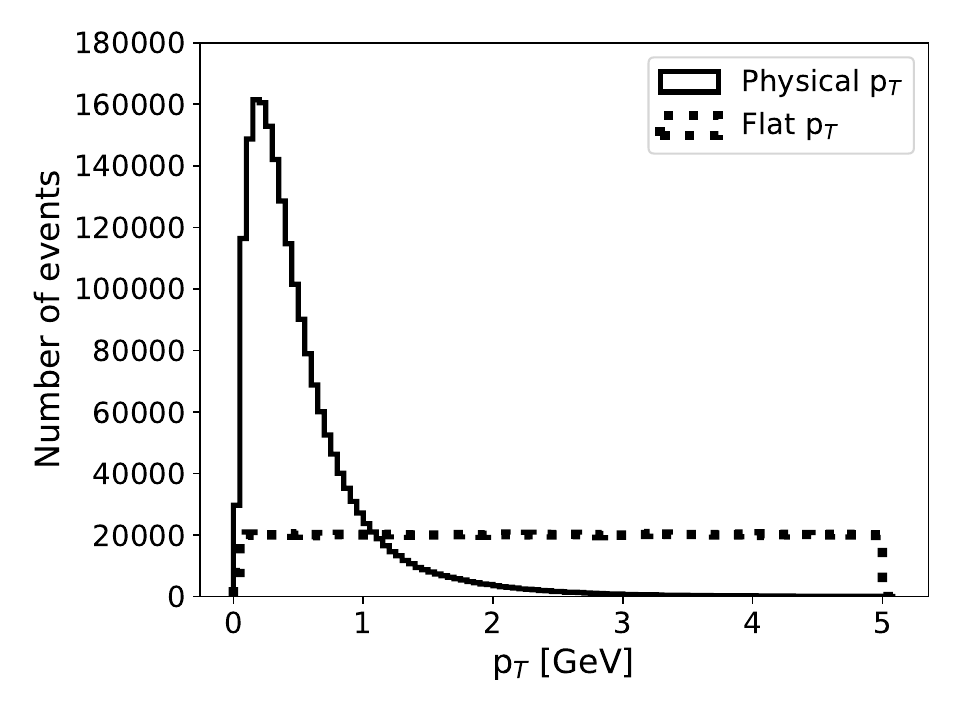}
    \includegraphics[width=0.32\textwidth]{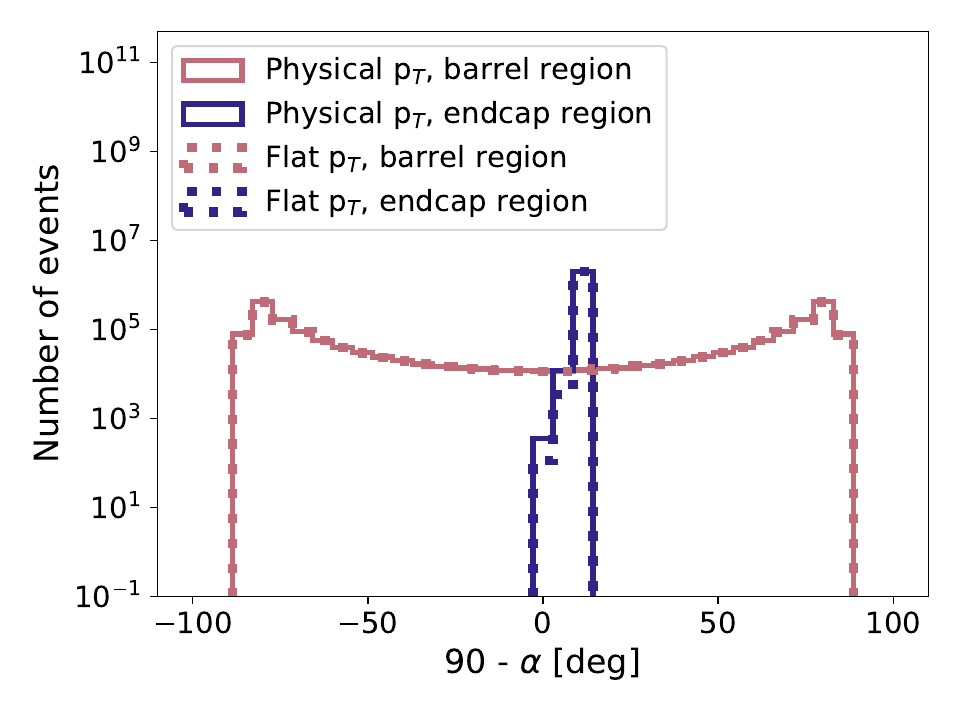}
    \includegraphics[width=0.32\textwidth]{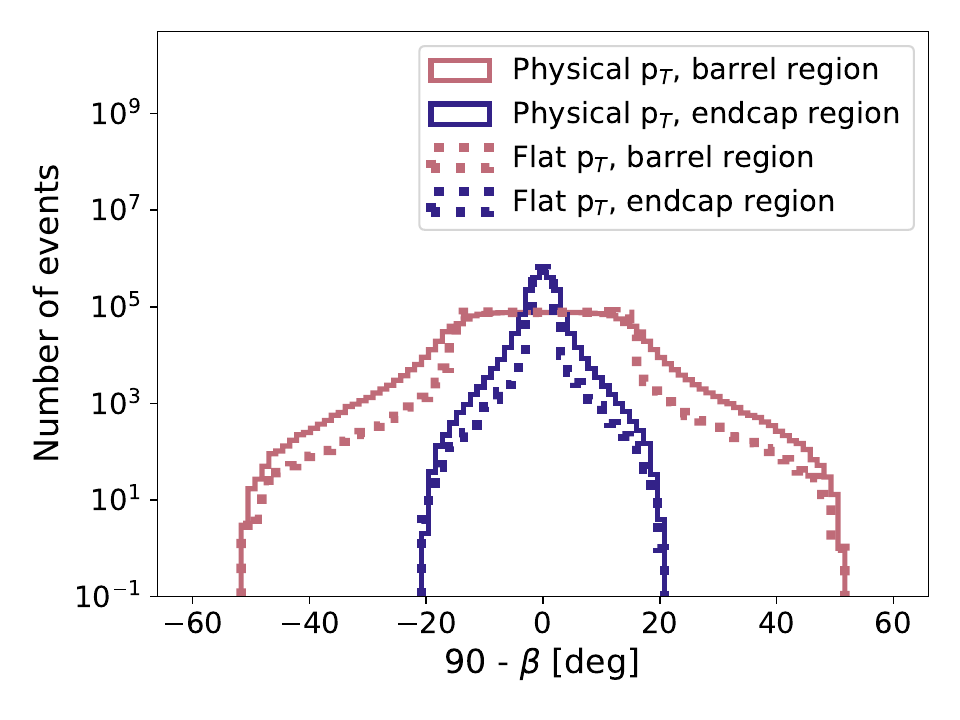}
    \caption{Distributions of p$_T$ (left), $\alpha$ (center), and $\beta$ (right) across the various datasets produced. For both angular variables, $\alpha = 90^\circ$ and $\beta = 90^\circ$ correspond to normal incidence. The p$_T$ distributions (left plot) for the physical and flat samples are identical for sensors in both the barrel and endcap layers. The distribution of $\alpha$ values in the endcap regions (center plot) is not symmetric because only one endcap disk was simulated.}
    \label{fig:truth-plots}
\end{figure}

Datasets with different p$_T$ distributions were also created for each geometry~\cite{zenodo_sp_datasets}.  One, called the ``flat-p$_T$" dataset, contains 2 million events with their p$_T$ distributed uniformly in [0.03, 5]~GeV.  This dataset is used for training in order to improve performance across a large momentum range and avoid bias. The incident position is uniformly distributed in $x$ and $y$ on each module, so is approximately uniformly distributed in global $z$ and $\phi$ for the barrel, and in global $R$ and $\phi$ for the endcap, resulting in the angular ($\alpha$ and $\beta$) distributions seen in Figure~\ref{fig:truth-plots}. The range of incident angles in the training datasets corresponds to the full range that would be seen over one layer of the barrel or one ring in a disk, so the size of this dataset is indicative of what would be required to train ASICs for one full layer of the detector. 

A second dataset, the ``physical-p$_T$'' dataset, follows a falling momentum distribution typical of what would be seen in the 13 TeV $pp$ collision data at CMS. This p$_T$ distribution was corrected for losses due to the inefficiency of the CMS tracker at very low p$_T$~\cite{Yoo_2024}. The distributions of p$_T$ observed for modules in the barrel and endcap layers were found to be similar; therefore, the same p$_T$ distribution was used to study both detector regions. The range of incident angles is not significantly different from the flat-p$_T$ dataset. These datasets, containing about 2 million events each, were used to test the performance of the network with a realistic p$_T$ distribution.

The distributions of p$_T$, $\alpha$, and $\beta$ from all events in both physical- and flat-p$_T$ datasets are shown in Figure~\ref{fig:truth-plots}.

\begin{figure}[tbh]
    \centering
    \includegraphics[width=0.4\linewidth]{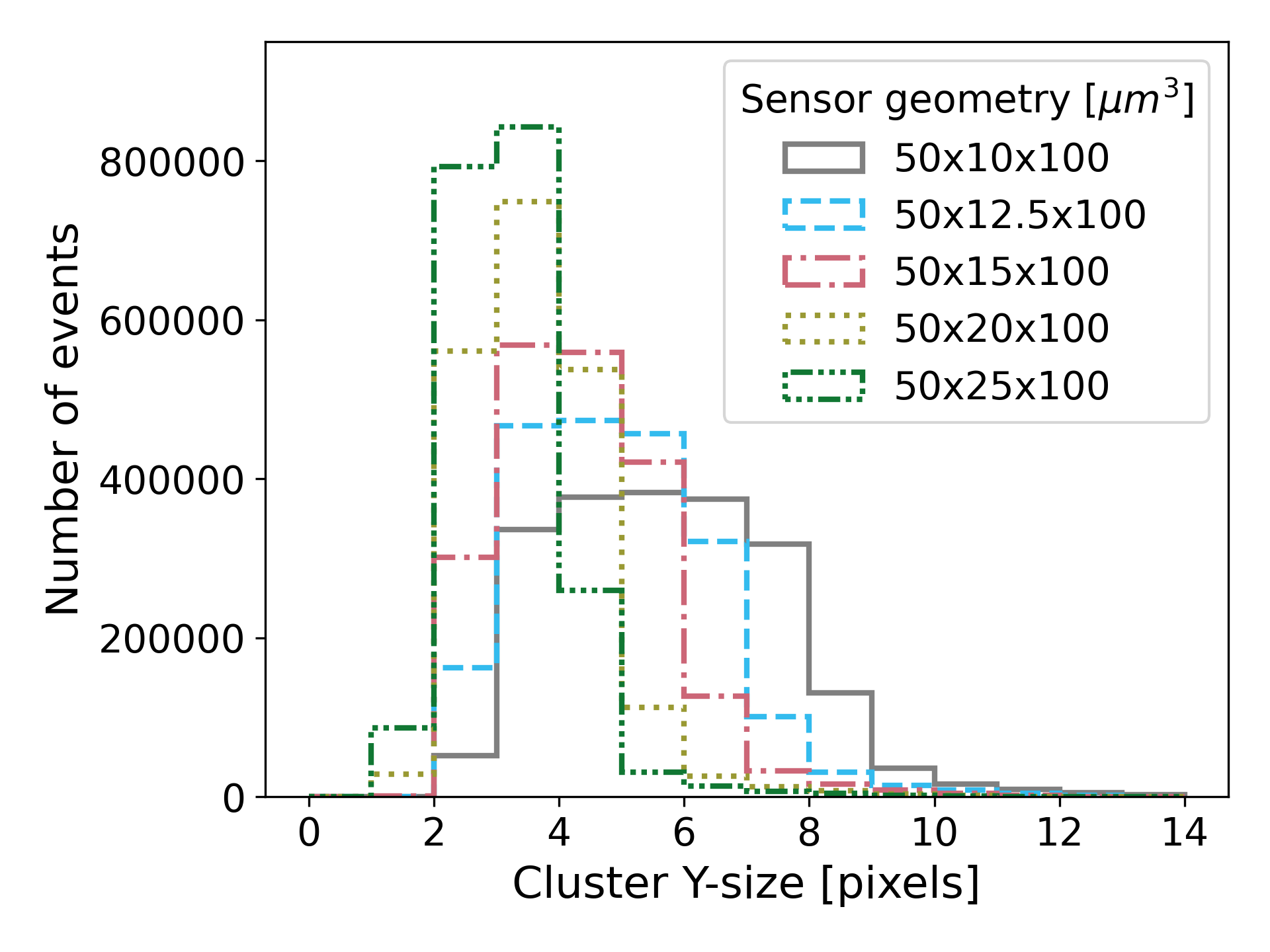}
    \includegraphics[width=0.4\textwidth]{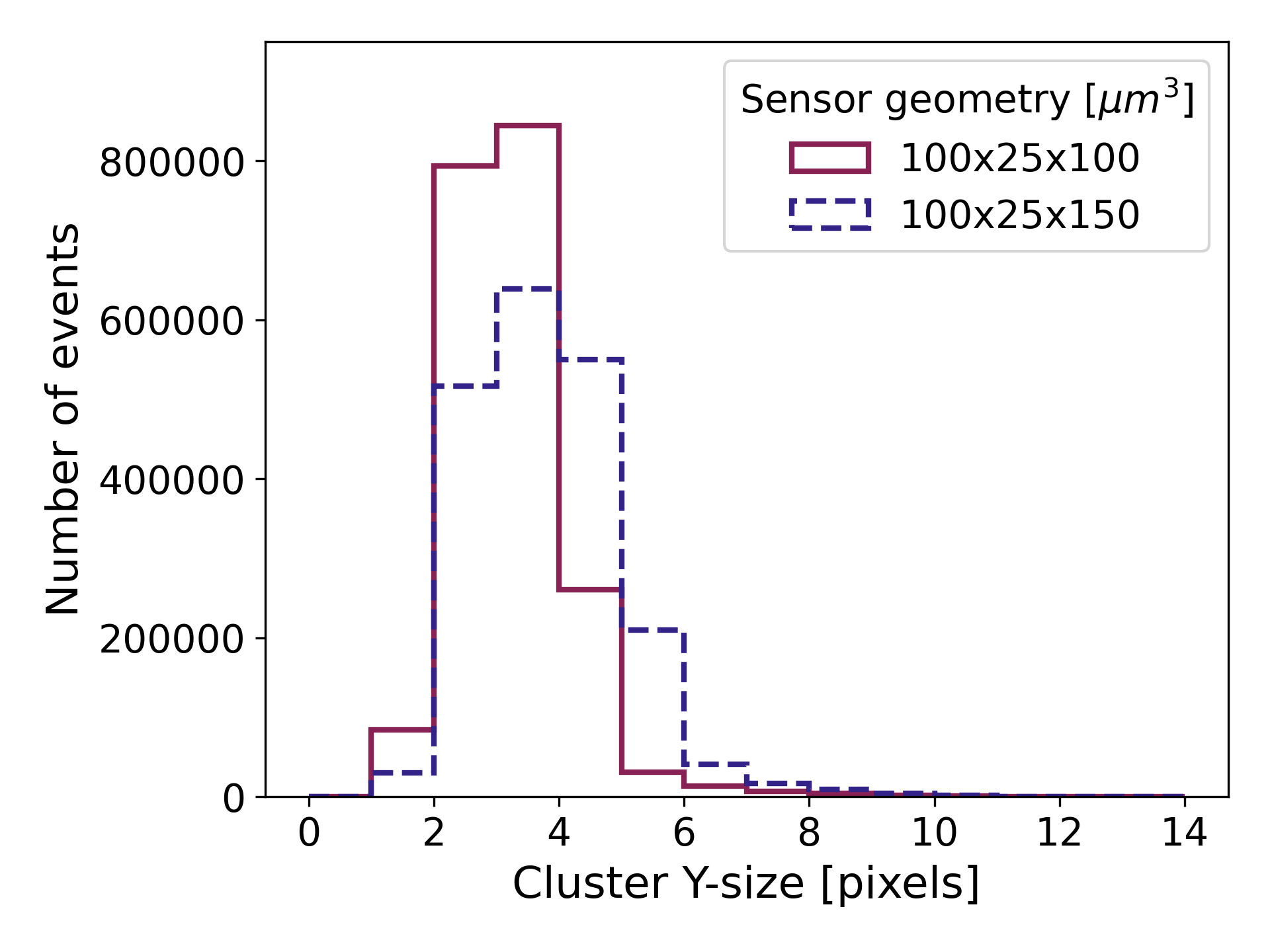}
    \caption{Distribution of cluster size along the $y$-direction for sensor geometries in the barrel region, using the physical-p$_T$ dataset. The distributions in left and right correspond to sensors with varying $y$-pitch and thickness, respectively.}
    \label{fig:cluster-y-size-distrib}
\end{figure}

The change in cluster size for different sensor geometries is visible in Fig.~\ref{fig:cluster-y-size-distrib}, which shows the distribution of cluster sizes along the $y$-direction for the physical-p$_T$ datasets of all sensor geometries studied.  

Pixel detectors are typically installed closest to the interaction point, and face the brunt of radiation damage.  This alters sensor characteristics, so an assessment of the evolution of the classifier performance with radiation damage is crucial. Therefore, three datasets corresponding to integrated luminosities of 0, 370, and 1100 fb$^{-1}$ of $pp$ collisions at center-of-mass energy of 13 TeV (equivalent to 0, 3.3$\times$10$^{15}$, and 1.0$\times$10$^{16}$ 1 MeV equivalent neutrons per \unit{\centi\meter}$^2$, respectively) were produced for the baseline (50$\times$12.5$\times$100 $\unit{\micro\meter}^3$) and CMS Phase-2 sensor geometries. A per-pixel charge threshold of 400 electrons was applied to remove induced signals from trapped charges. The datasets from the unirradiated sensors however had no charge threshold as noise has not been introduced in these studies yet.

Table~\ref{tab:sim-datasets} summarizes all datasets produced to obtain the results presented in this paper. The bias voltages mentioned in this table ensure similar electric field values in the bulk region for the unirradiated sensors. The bias was increased in case of radiation damaged sensors to help retain optimal electric field magnitudes.

\begin{table}[h]
\centering
\resizebox{4in}{!}{%
\begin{tabular}{|c|c|c|c|c|}
\hline
Sl. No. &
  \begin{tabular}[c]{@{}c@{}}Detector \\ region\end{tabular} &
  \begin{tabular}[c]{@{}c@{}}Sensors \\ simulated\end{tabular} &
  \begin{tabular}[c]{@{}c@{}}Angle between\\ \textbf{E} and \textbf{B}\end{tabular} &
  \begin{tabular}[c]{@{}c@{}}Irradiation level\\ {[}fb$^{-1}${]}\end{tabular} \\ \hline
1 & Barrel & S1 - S7 & 90$^\circ$ & 0         \\ \hline
2 & Endcap & S1 - S7 & 180$^\circ$ & 0         \\ \hline
3 & Barrel & S1, S7  & 90$^\circ$ & 370, 1100 \\ \hline
\end{tabular}%
}
\caption{Categorization of the simulation datasets based on their use-cases in studies presented in this paper.}
\label{tab:sim-datasets}
\end{table}

\section{Filtering algorithm for pixel clusters}\label{sec:filter}

Three classification models were evaluated in the baseline study~\cite{Yoo_2024} and the model chosen for deployment on-sensor is the one used for the studies presented in this paper. 
The classifier is trained using the ``cluster $y$-profile'' of each event, which contains more information than the cluster size alone.  The cluster $y$-profile is calculated for each event as the sensor output at \mbox{4000~\unit{\pico\second}}  summed over pixel rows ($x$) to project the integrated charge of the cluster onto the $y$-axis. Additionally, $y_0$, the azimuthal position of the particle's incident position on the sensor array in the local detector coordinates, is an input to the network and for the work in this paper the truth value is used. In future iterations, it can be extracted from the coordinates of the pixel cluster, and a precision at the millimeter scale is sufficient for the p$_T$ classification task. The inter-dependence of the cluster $y$-profile, $y_0$, and p$_T$ was examined in~\cite{Yoo_2024}. This network, therefore taking in 14 input values ($y_0$ and 13 values from cluster $y$-profile), contains one dense hidden layer with 128 neurons and 1920 parameters, followed by one dense output layer with 3 neurons and 387 parameters. A softmax activation is used to generate classification probabilities between 0 and 1, and each event is assigned the classification label corresponding to the highest probability. 

The three output categories correspond to low p$_T$ oppositely charged particle tracks and high p$_T$ tracks (independent of charge). The rationale to have two background categories comes from differences in cluster profiles for oppositely charged particle tracks that curve in opposite directions in the magnetic field, and is discussed in more detail in \cite{Yoo_2024}. During training, tracks are classified as high or low p$_T$ depending on whether their transverse momentum is above or below a threshold $X$, referred to as the ``p$_T$ boundary.’’ In previous studies, $X$~=~0.2~GeV was chosen, as it yielded a flat signal efficiency for tracks with p$_T$~\textgreater~2~GeV, which is a useful range for physics studies~\cite{Yoo_2024}. In this work, we re-evaluate the performance for p$_T$ boundary values in the range 0.1–0.5~GeV.

Neural network trainings were run for 200 epochs and stopped early if the loss function showed no improvement after 20 epochs. A batch size of 1024 was used in all models. The Adam optimizer with a learning rate of 0.001 was used in conjunction with the Keras Sparse Categorical Cross entropy loss function in all models.

\subsection{Performance metrics}
The model performance is evaluated using three metrics:
\begin{itemize}
\item Signal efficiency: fraction of clusters with p$_T$~\textgreater~2~GeV that are classified as high p$_T$.
\item Background rejection: fraction of clusters with p$_T$~\textless~2~GeV that are classified as low p$_T$.
\item Data-reduction: fraction of clusters classified as low p$_T$ irrespective of true class
\end{itemize}
We calculate a statistical uncertainty based on the dataset size using the binomial uncertainty formula. All training and evaluation cycles were performed 10 times with different random seeds to obtain stochastic errors from the ML algorithm. The final uncertainty is defined as the addition of the statistical and stochastic errors in quadrature.

\section{Studies on performance of filtering algorithm}\label{sec:results}

The following sub-sections describe the findings of the studies in terms of the performance metrics defined in Section~\ref{sec:filter}. We do not consider untracked clusters and single-pixel clusters; therefore, the data-reduction results presented in this paper are understood to be an underestimate~\cite{Yoo_2024}. 

\subsection{Sensor geometry}\label{sec:geometry}
\begin{figure}[H]
    \centering
    \includegraphics[width=0.32\textwidth]{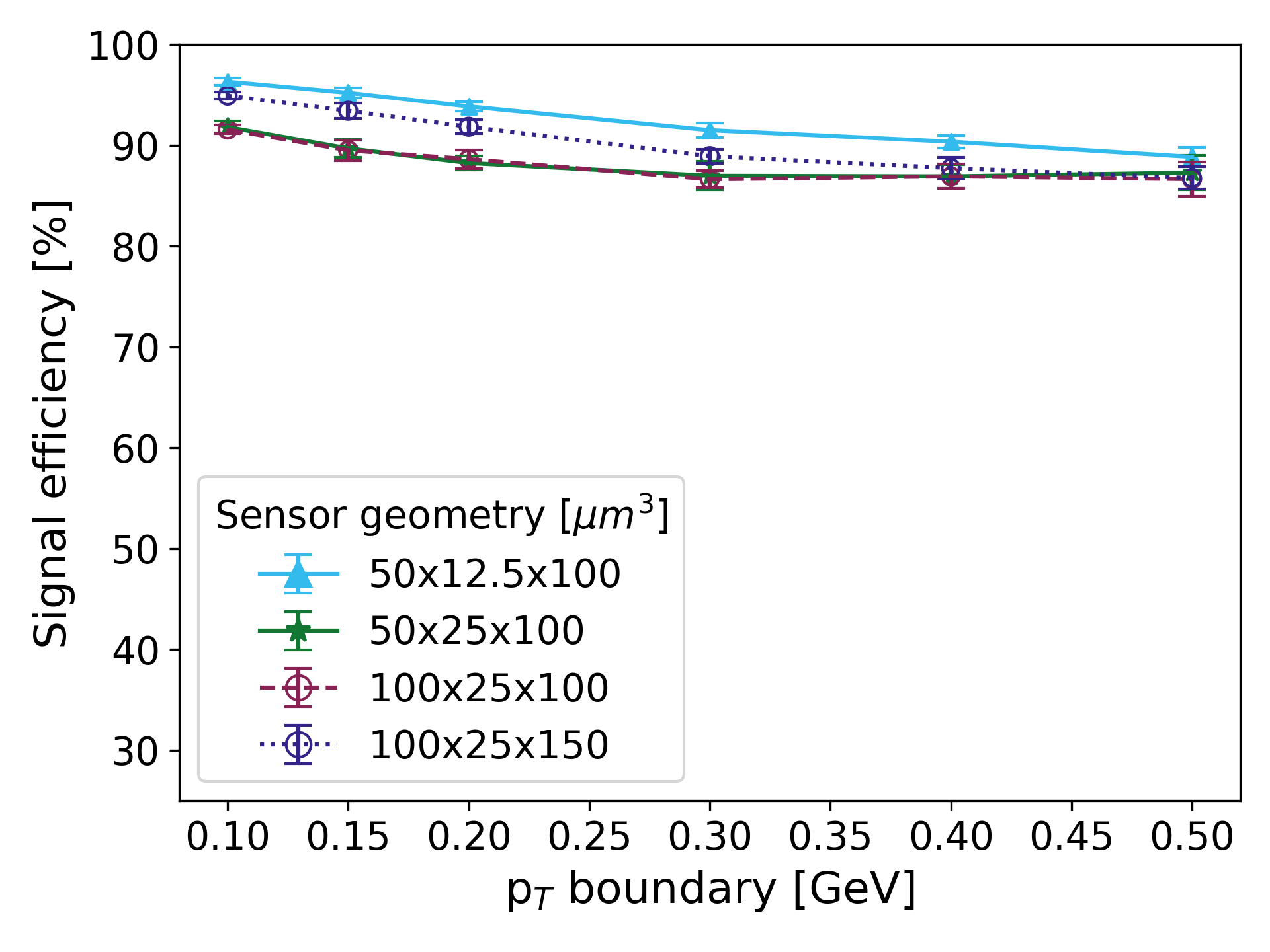}
    \includegraphics[width=0.32\textwidth]{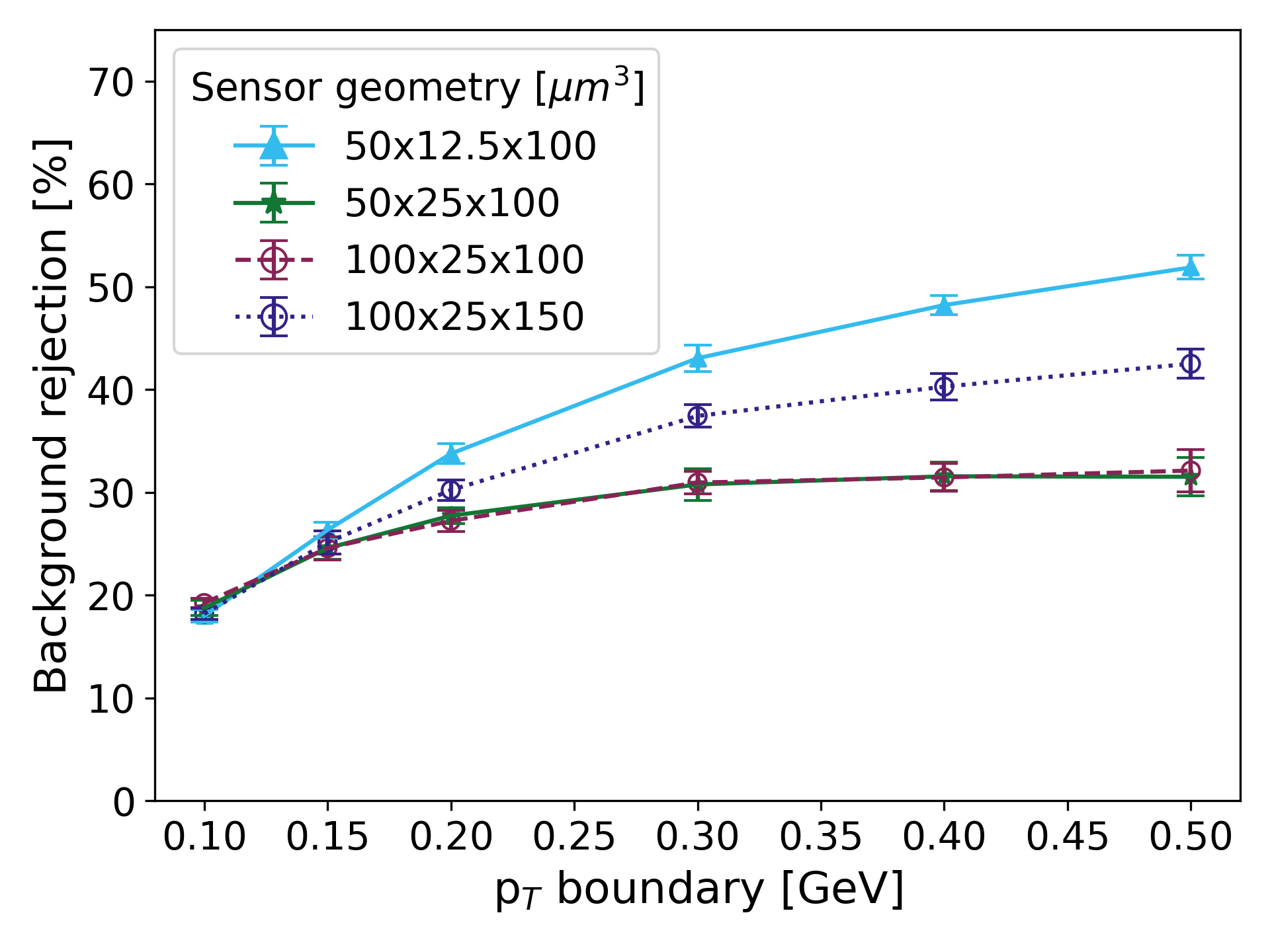}
    \includegraphics[width=0.32\textwidth]{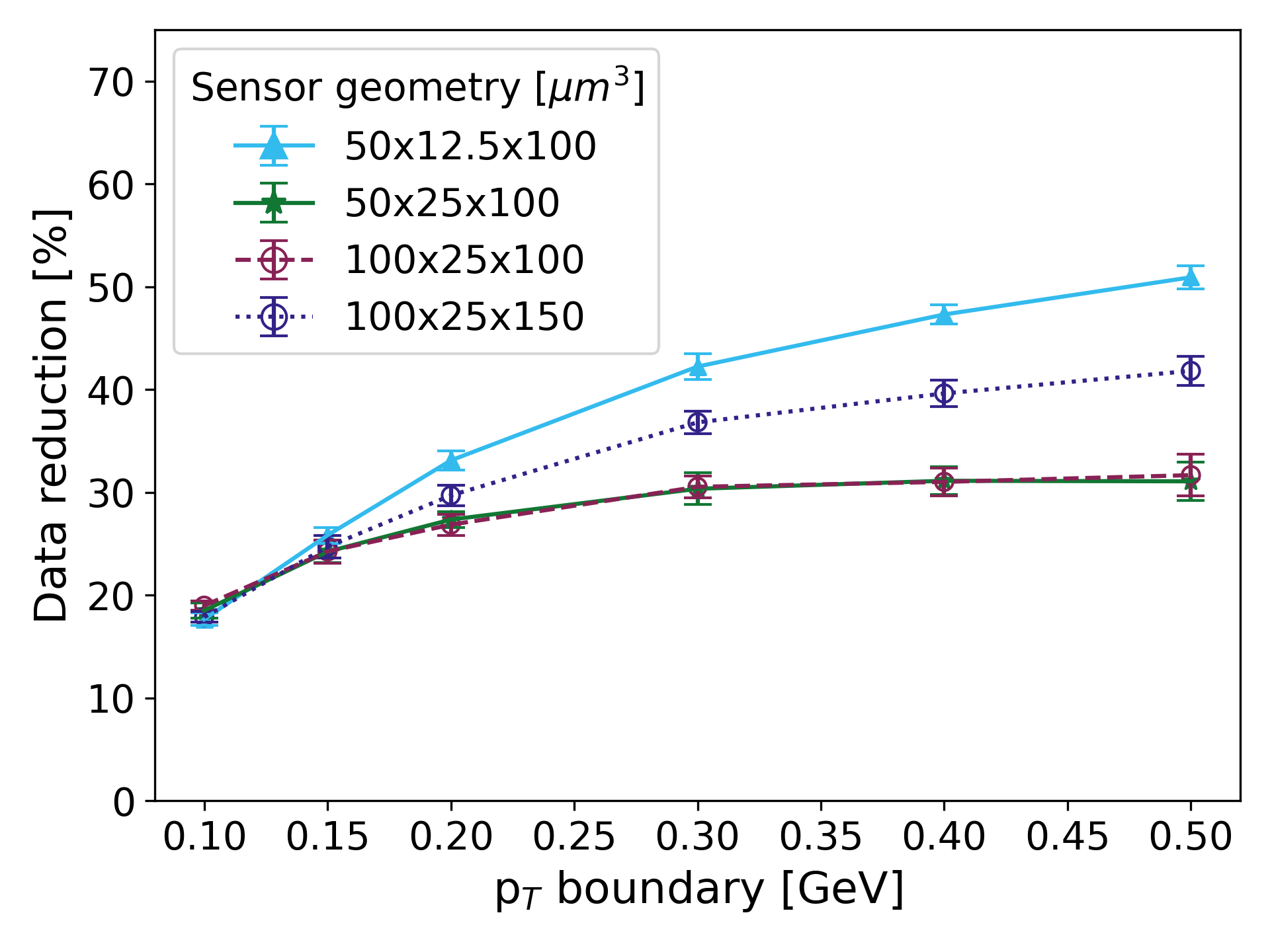}
    \caption{Performance results of sensors in the barrel region for four sensor geometries as a function of p$_T$ boundary. The legend describes the pixel dimensions along the $x$ (length), $y$ (width), and $z$ (thickness) directions, respectively.}
    \label{fig:sensor-geometry-study-allpt}
\end{figure}

The impact of sensor geometry on model performance was studied using datasets from seven configurations, varying in both pitch and thickness. The performance of the algorithm trained on the flat-p$_T$ and evaluated on the physical-p$_T$ datasets is presented in Fig.~\ref{fig:sensor-geometry-study-allpt}. The performance is plotted as a function of p$_T$ boundary for four sensor geometries. We observe a performance drop across all p$_T$ boundaries for increasing $y$-pitch and decreasing thickness. 

\begin{figure}[H]
    \centering
    \includegraphics[width=0.32\textwidth]{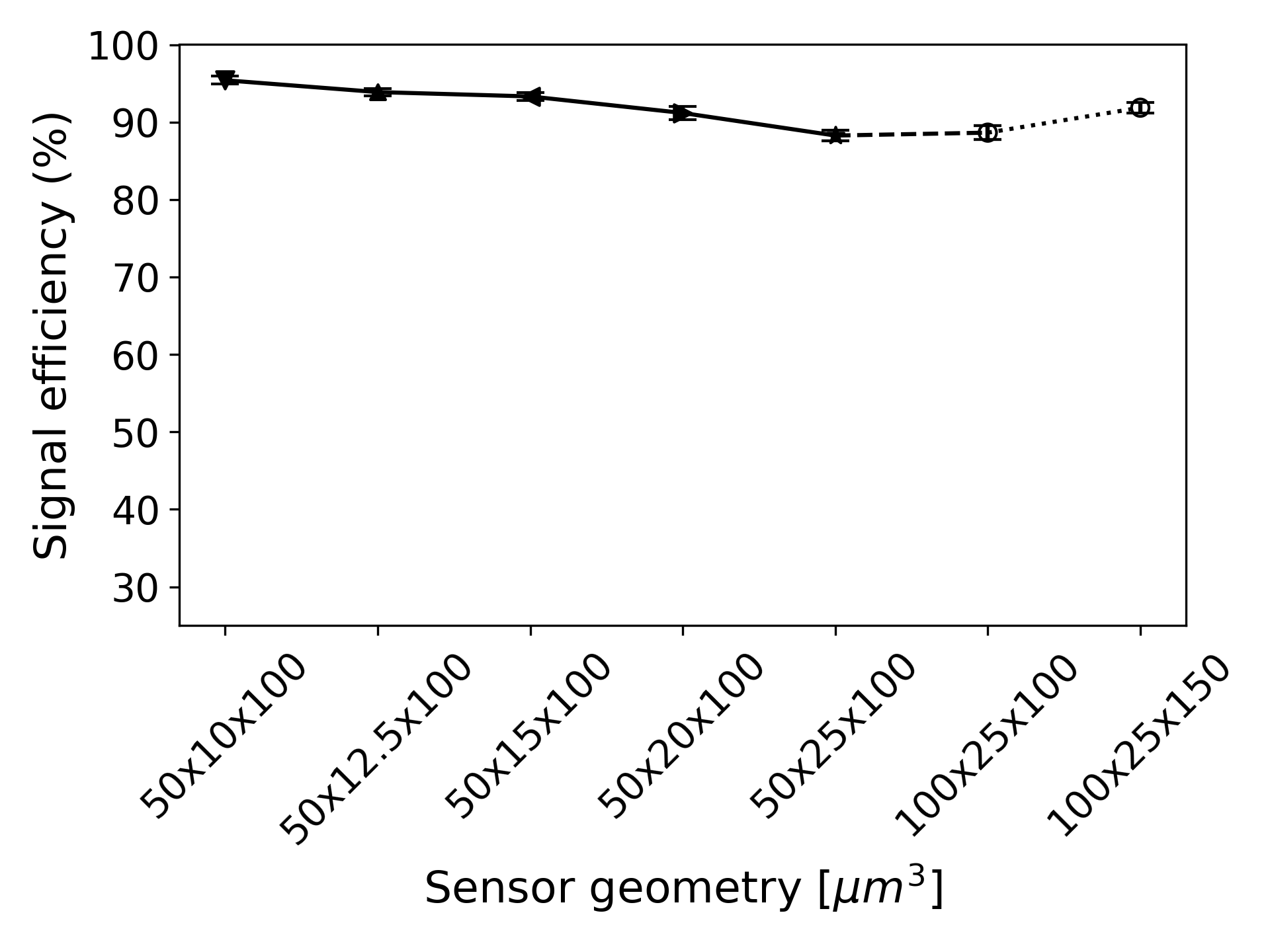}
    \includegraphics[width=0.32\textwidth]{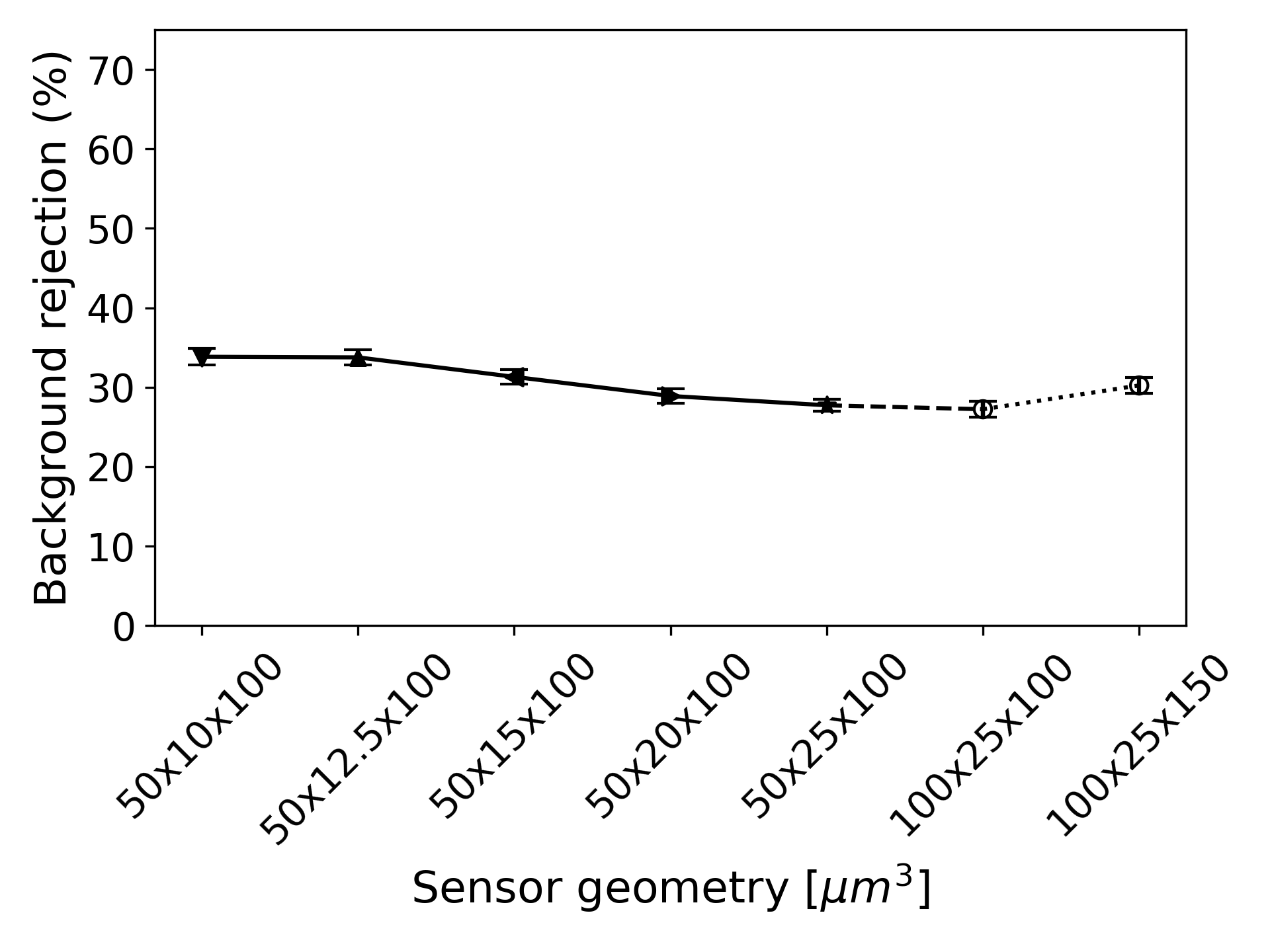}
    \includegraphics[width=0.32\textwidth]{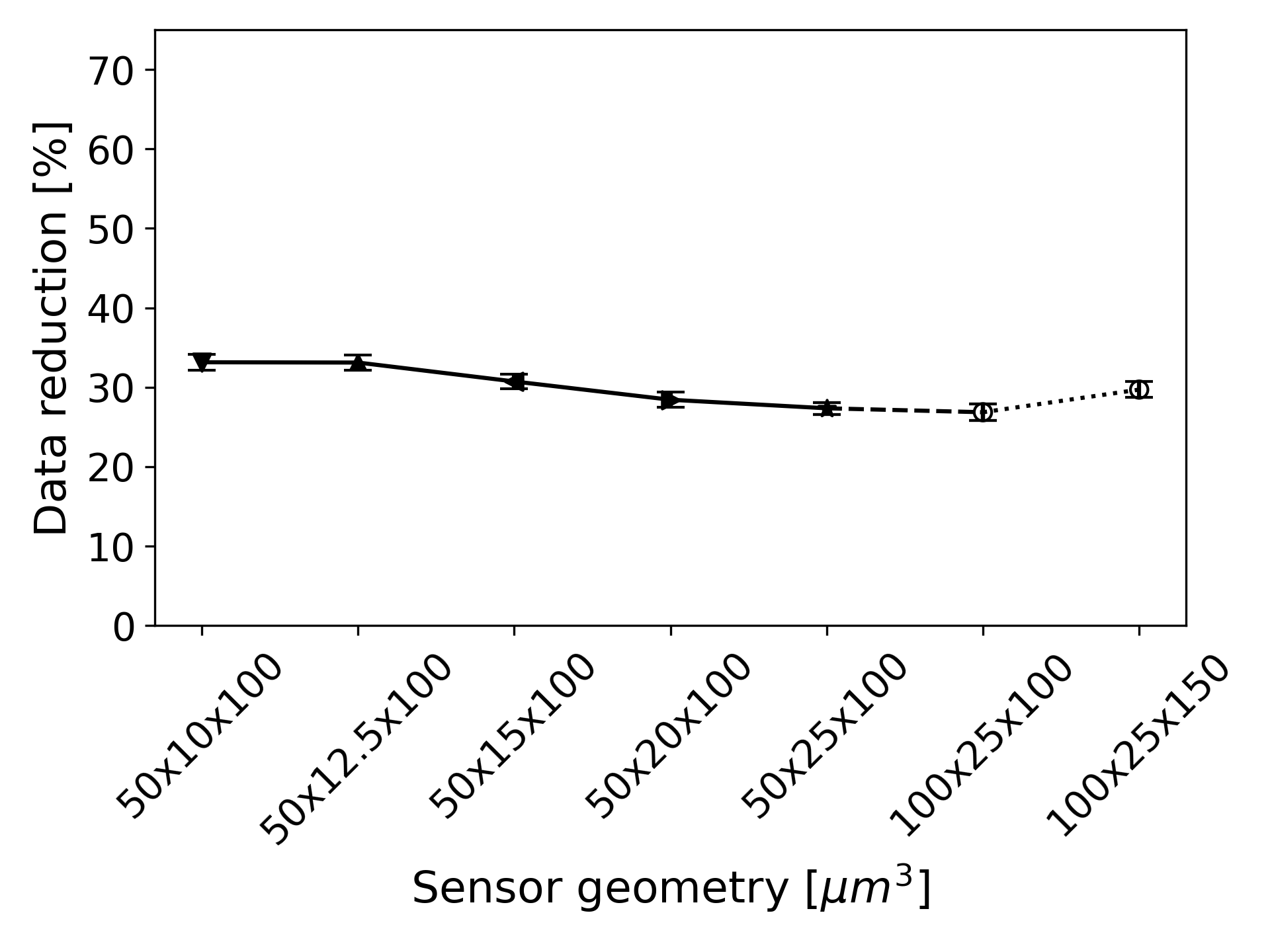}
    \caption{Performance results of sensors in barrel region for sensor geometries varying in pitch and thickness. The $x$-axis labels indicate unit pixel dimensions along the $x$ (length), $y$ (width), and $z$ (thickness) directions, respectively. The solid, dashed, and dotted lines correspond to geometries with varying $y$-pitch, $x$-pitch, and thickness, respectively.}
    \label{fig:sensor-geometry-study}
\end{figure}

The hyperparameter optimization study carried out in prior work~\cite{Yoo_2024} for the 50$\times$12.5$\times$100 $\unit{\micro\meter}^3$ sensor geometry found the p$_T$ boundary value of 0.2~GeV to be a reasonable baseline value. We observe the same value produces acceptable performance across other sensor geometries as well. Results shown in Fig.~\ref{fig:sensor-geometry-study} correspond to model performance across all sensor geometries with the p$_T$ boundary hyperparameter set to 0.2~GeV.

As seen in Figure~\ref{fig:cluster-y-size-distrib}, the discriminating power on cluster profile features is lost with increasing sensor $y$-pitch. In line with this observation, we observe all three performance metrics to degrade as sensor $y$-pitch increases, and improve for increases in sensor thickness. The performances of the 50$\times$25$\times$100 and 100$\times$25$\times$100 $\unit{\micro\meter}^3$ sensors are comparable within error margins, since the model input does not include any $x$-direction-dependent information.  Notably, the increase in thickness from 100$\times$25$\times$100 $\unit{\micro\meter}^3$ to 100$\times$25$\times$150 $\unit{\micro\meter}^3$ recovers some of the performance loss from the larger pixel pitch, though it does not perform as well as the 50$\times$12.5$\times$100 $\unit{\micro\meter}^3$ baseline.  

To assess the dependence on the per-pixel charge threshold, the network performance was evaluated for both the baseline and CMS Phase-2 sensor geometries over a threshold range of 0\textendash{}1500 electrons, consistent with values used in current detector systems. We observe that increasing the threshold up to 1500 electrons and retraining the network (at 0.2~GeV p$_T$ boundary) results in a decrease in performance of less than 5 percentage points compared to the zero-threshold case for both sensor geometries.

\subsection{Lorentz drift and barrel versus endcap}\label{sec:lorentz}
\begin{figure}[H]
    \centering
    \includegraphics[width=0.32\textwidth]{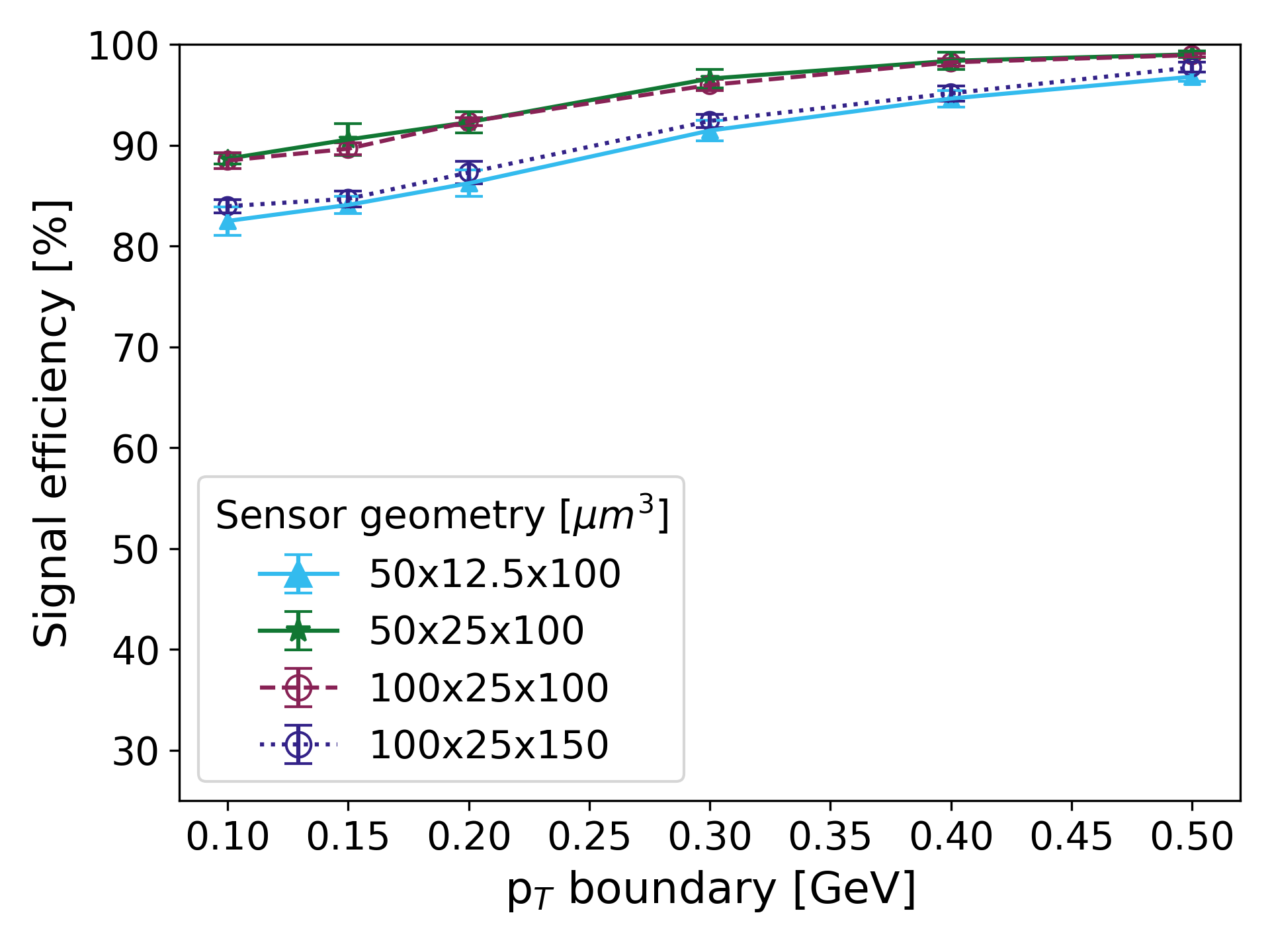}
    \includegraphics[width=0.32\textwidth]{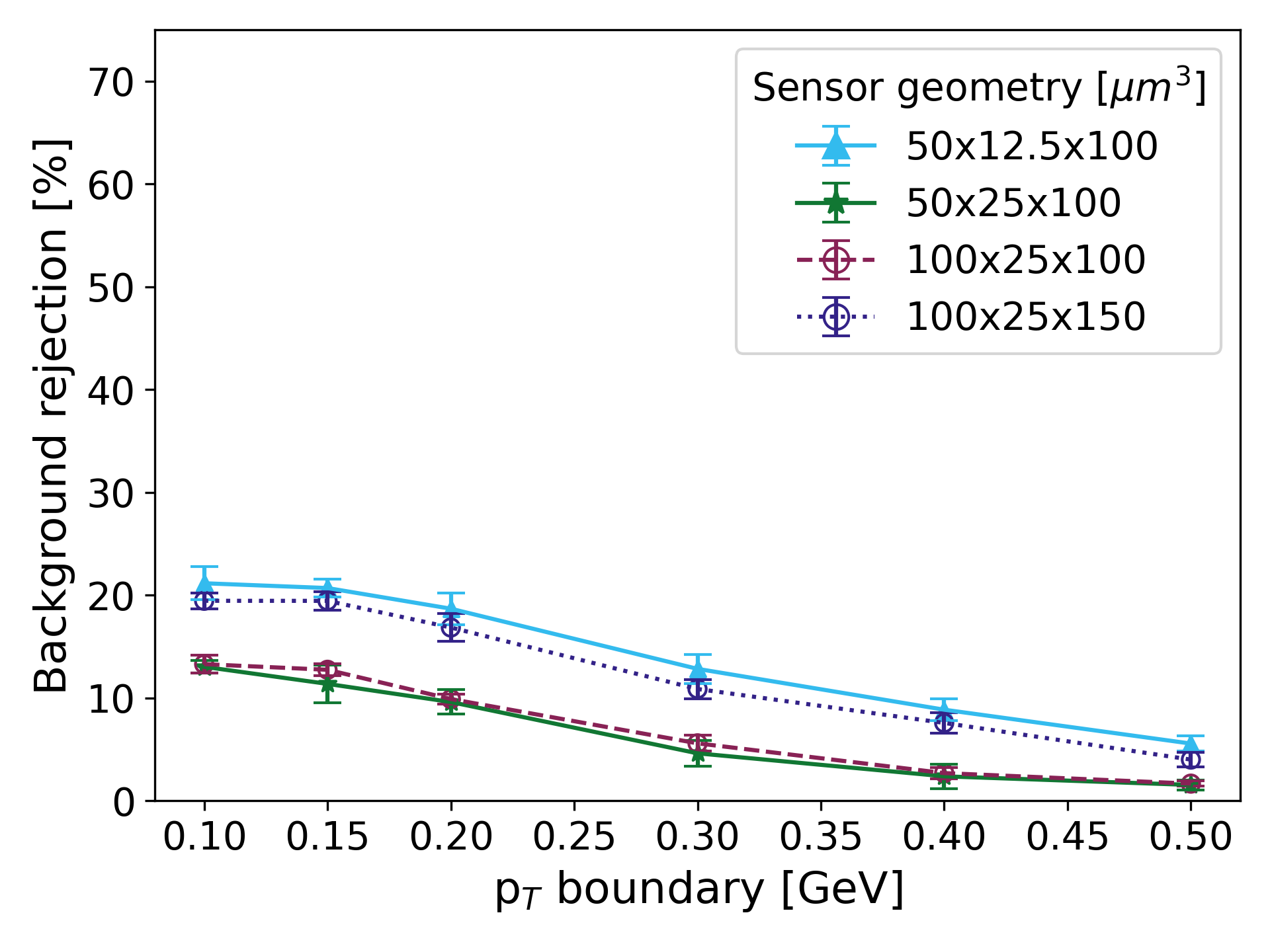}
    \includegraphics[width=0.32\textwidth]{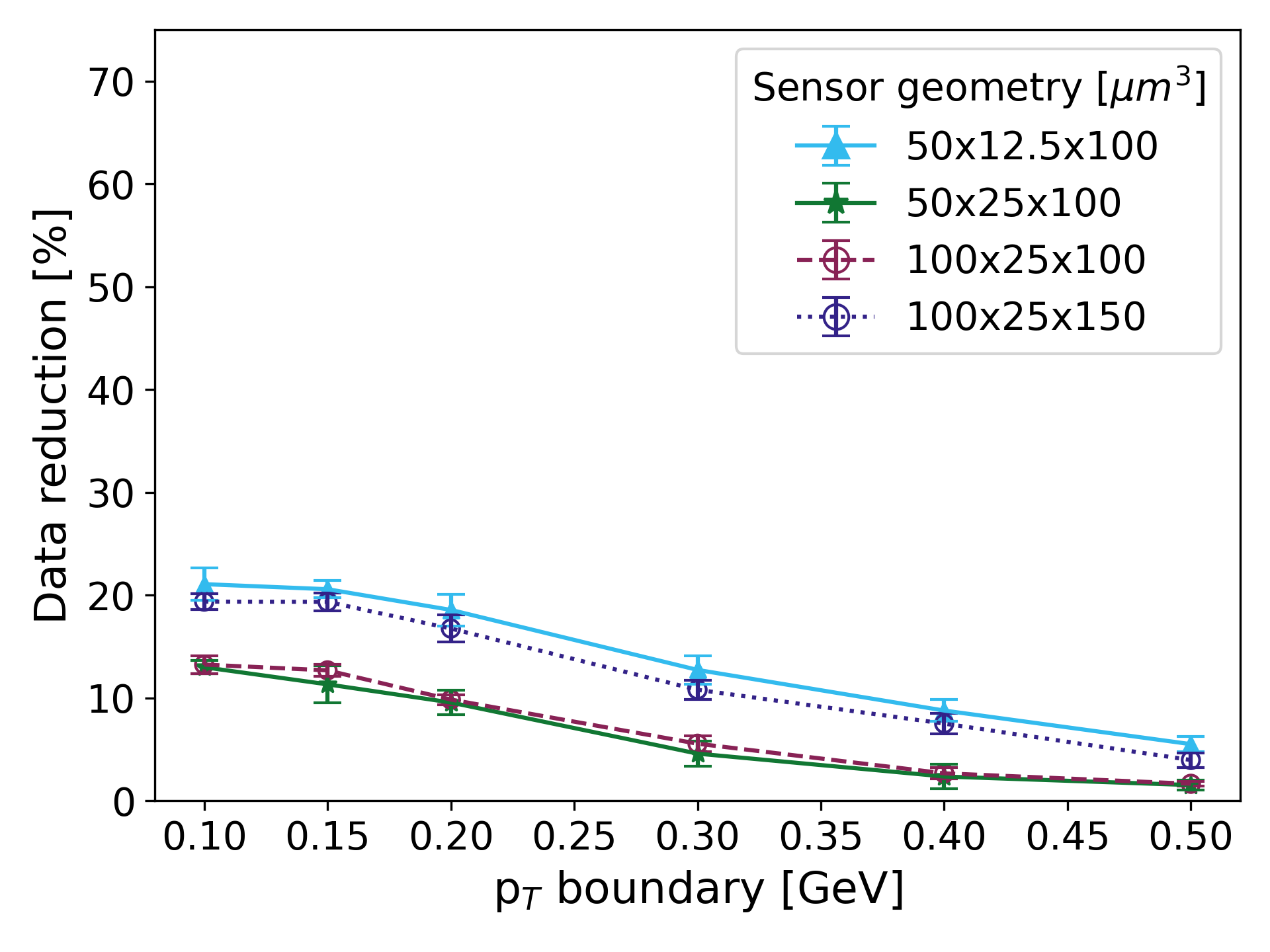}
    \caption{Performance results of sensors in the endcap region for four sensor geometries as a function of p$_T$ boundary. The legend describes the pixel dimensions along the $x$ (length), $y$ (width), and $z$ (thickness) directions, respectively.}
    \label{fig:no-lorentz-drift-study}
\end{figure}

Model performance was also evaluated in detector regions with no Lorentz drift, where the electric field within the sensor is parallel to the external magnetic field.  This is typical of disk-type (endcap) geometries in conventional pixel detectors when planar-sensor modules are oriented orthogonal to the magnetic field.  Figure~\ref{fig:no-lorentz-drift-study} shows these results as a function of the p$_T$ boundary for four sensor geometries.

The trend of the performance versus p$_T$ boundary for sensors in the endcap contrasts with that of sensors in the barrel: with increasing p$_T$ boundary, signal efficiency improves while data reduction and background rejection decrease, for all geometries. At the 0.2~GeV boundary, sensors in the endcap region show reduced performance relative to those in the barrel region.

The observed behavior in the results presented above reflects both the absence of Lorentz drift and the distinct kinematic distributions of charged particles reaching the endcap regions. To isolate the effect of Lorentz drift, dedicated PixelAV simulations were performed for barrel-layer pixels with the magnetic field effectively set to zero within the silicon volume, relevant for 3D sensors where Lorentz drift is suppressed. In this configuration, the network performance at a 0.2~GeV p$_T$ boundary is reduced by less than 6\% in absolute terms compared to the nominal case across all sensor geometries and all three performance metrics.

A complementary study was carried out to evaluate the impact of the external magnetic field strength on network performance for the baseline sensor geometry in the barrel-layer. Using datasets for magnetic field configurations of 3.8~T and 2~T, and the neural network retrained independently for each scenario, we observed the network performance to worsen with the reduction in magnetic field strength. At the 0.2~GeV p$_T$ boundary, the performance in the 2~T configuration is lower by less than 8\% in absolute terms compared to the 3.8~T case across all three performance metrics.

\subsection{Irradiation}\label{sec:irradiation}
Using datasets produced for radiation damaged sensors for the baseline and CMS Phase-2 sensor geometries, we examined the classification network's performance and the importance of having reprogrammable weights in the ASIC.  

\begin{figure}[h]
    \centering
    \includegraphics[width=0.32\textwidth]{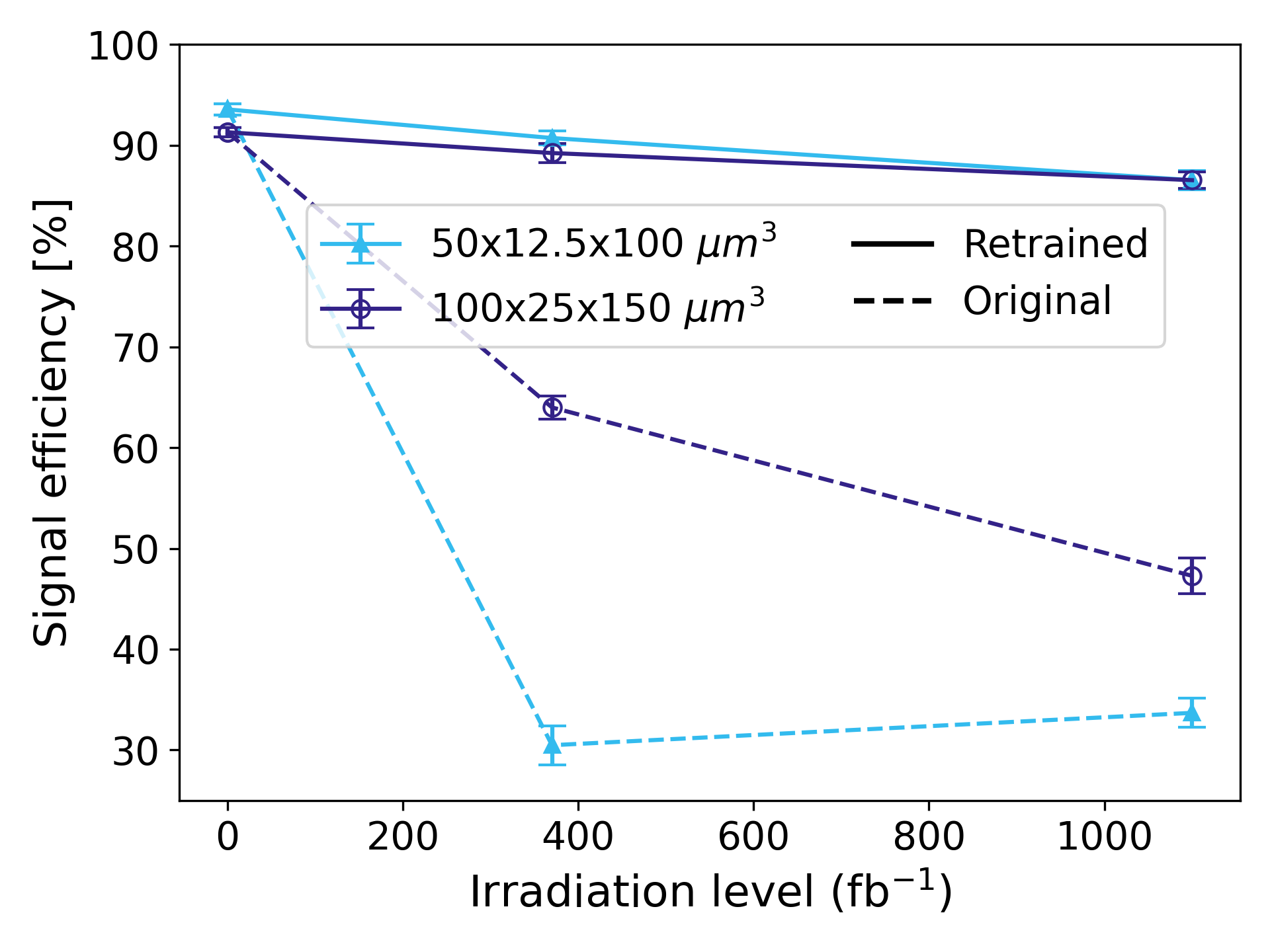}
    \includegraphics[width=0.32\textwidth]{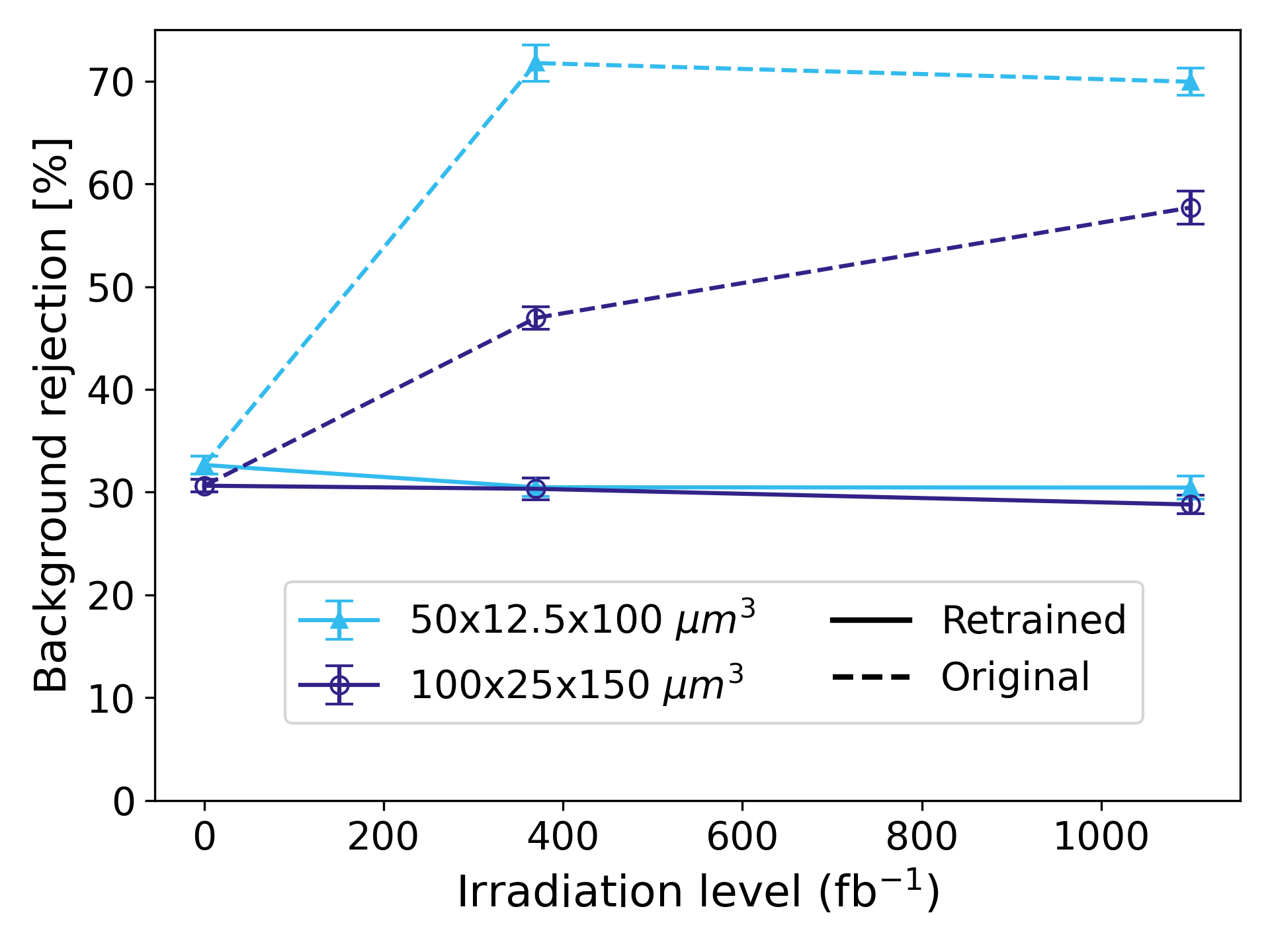}
    \includegraphics[width=0.32\textwidth]{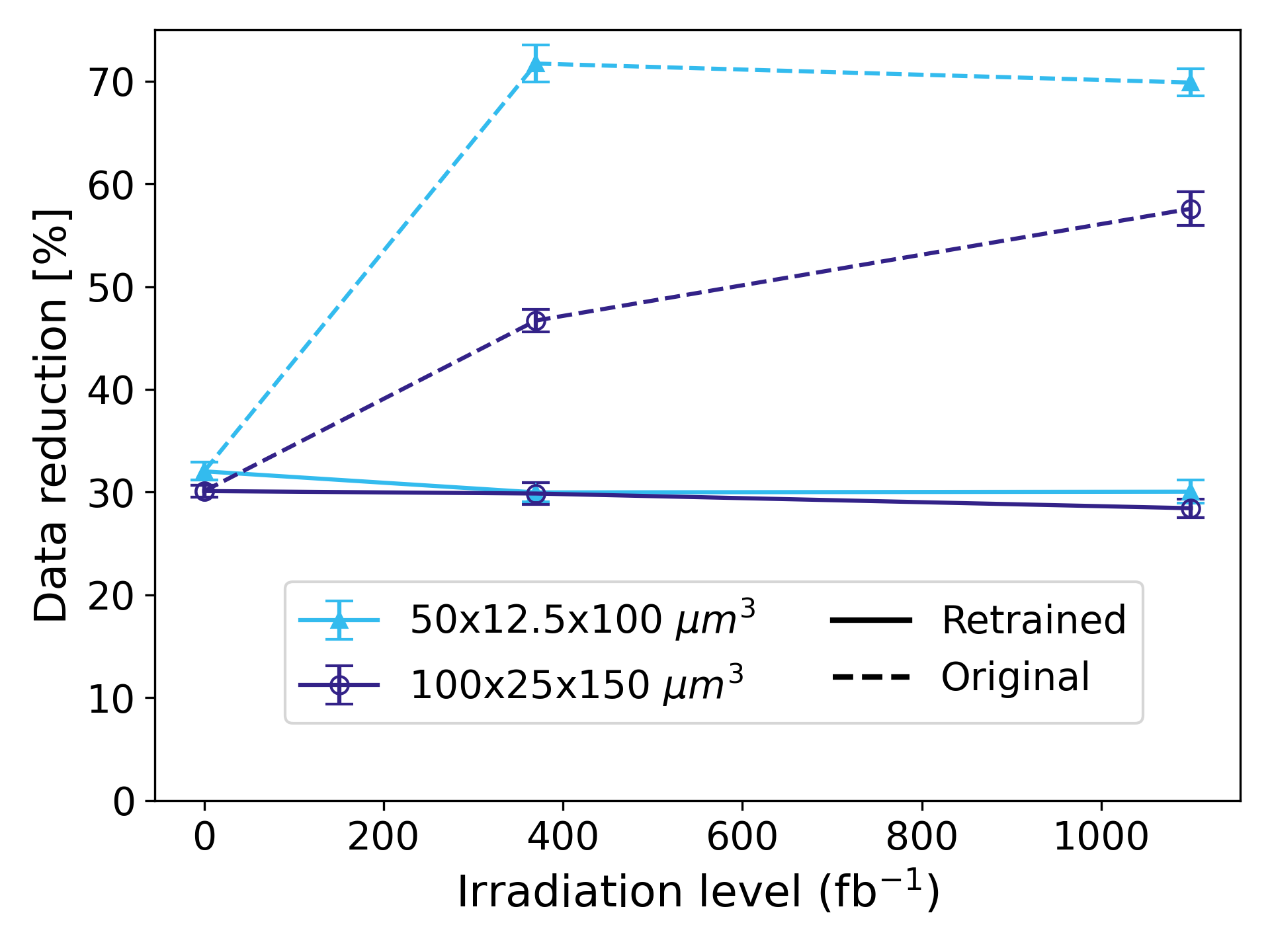}
    \caption{Performance results of the baseline and CMS Phase-2 sensors in the barrel region as a function of irradiation levels. ``Original" means using the weights resulting from training on simulation datasets of unirradiated detectors. Results in this figure use a $p_T$ boundary = 0.2~GeV with the networks trained and evaluated on datasets with a 400 electron per-pixel threshold to address leakage currents arising from trapped charges in irradiated sensors.}
    \label{fig:irradiation-study}
\end{figure}

As shown in Figure~\ref{fig:irradiation-study}, if the model weights remain fixed while radiation damage accumulates, classification accuracy deteriorates significantly\textemdash{}leading to high rejection rates even for high p$_T$ tracks. In contrast, retraining the models at each irradiation level (and subsequently updating ASIC weights) preserves model performance, demonstrating that the performance of the classifier can be mostly maintained even at fluences typical of the HL-LHC through simply updating the weights. For example, if retrained, the signal efficiency and data reduction of the baseline sensor geometry remain within 7\% and 2\% in absolute terms of their original values, respectively, across 1100 fb$^{-1}$ of radiation damage.

\subsection{Noise}\label{sec:noise}

To estimate the effect of detector noise on the performance of the filtering algorithm, a noise model is implemented by applying Gaussian smearing to the simulated charge collected in each pixel. The width of this Gaussian depends on the baseline charge fluctuation $Q_\text{base}$, the charge mismatch $Q_\text{th}$, and the Equivalent Noise Charge (ENC):
\begin{equation}
\sigma_\text{noise} = \sqrt{Q_\text{base}^2 + Q_\text{th}^2 + \text{ENC}^2}
\end{equation}
where $Q_\text{base}$ is the measure of the fluctuation in the pre-amplifier output baseline arising from event-by-event amplitude variation (due to variation in deposited charges by minimum-ionizing particles in Silicon) and variation in time between subsequent events. $Q_\text{th}$ represents the non-uniformity in the charge threshold across a pixel matrix that can arise from effects like systematic offset or random process variations in the digitization stage. ENC refers to the amount of charge that, if deposited in the sensor, would produce a signal equal to the root-mean-square value of the noise of the system.

Simulations of the architecture and ROIC process used for implementation of the $p_T$ filter \cite{roicpaper} were designed in the Cadence Virtuoso flow and simulated with Spectre \cite{cadence} in the baseline 50$\times$12.5$\times$100~$\unit{\micro\meter}^3$ sensor geometry. The resulting $Q_\text{base}$ is found to be 12 electrons, and $Q_\text{th}$ lies in the 50-60 electron range.  The ENC varies with sensor geometry and capacitance, temperature, and radiation damage. Simulations performed at -40$^\circ$C and sensor capacitance of 30 fF yield an ENC of 55 electrons before radiation damage. For a sensor leakage current of 1 nA, the same range of capacitance yields ENC of 65 electrons at -40$^\circ$C. This corresponds to the effect of leakage current compensation after irradiation to approximately 500 Mrad \cite{rd53specs}.

\begin{figure}[h]
    \centering
    \includegraphics[width=0.32\textwidth]{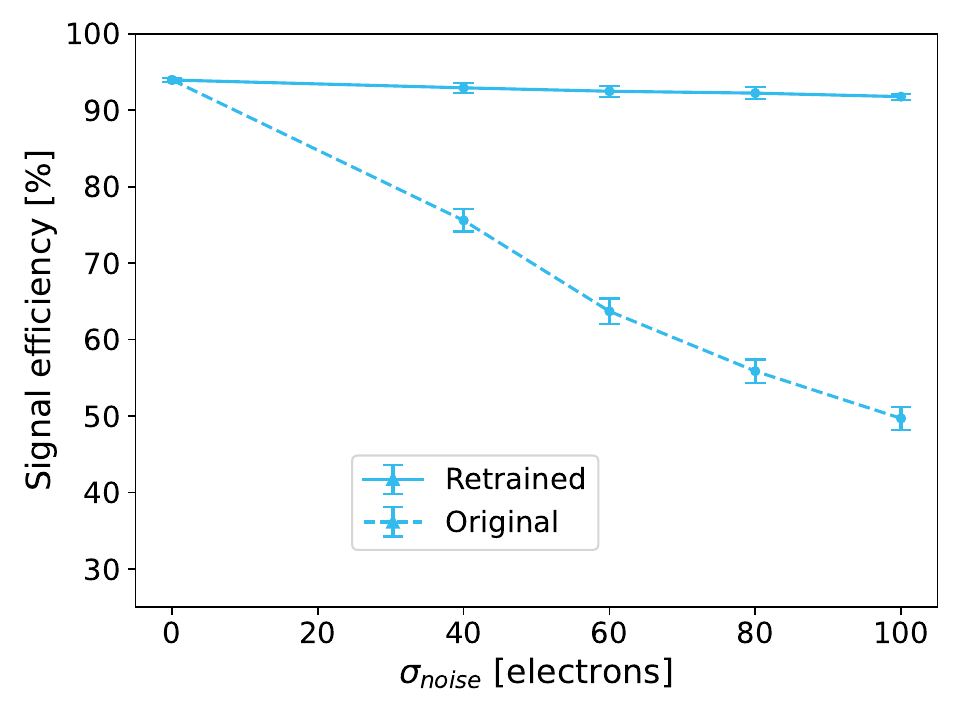}
    \includegraphics[width=0.32\textwidth]{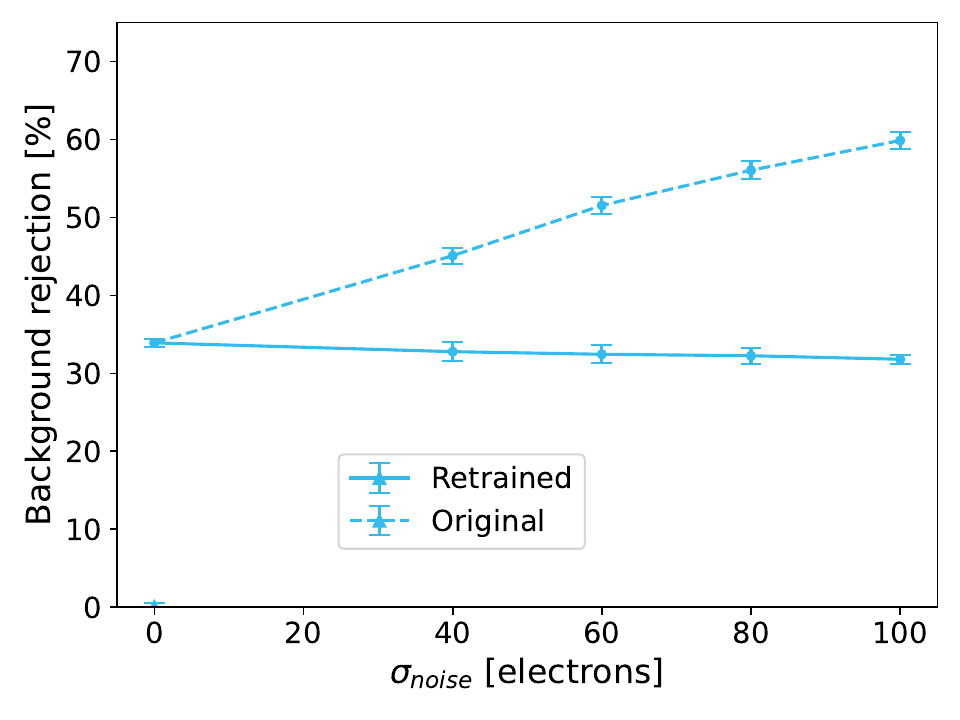}
    \includegraphics[width=0.32\textwidth]{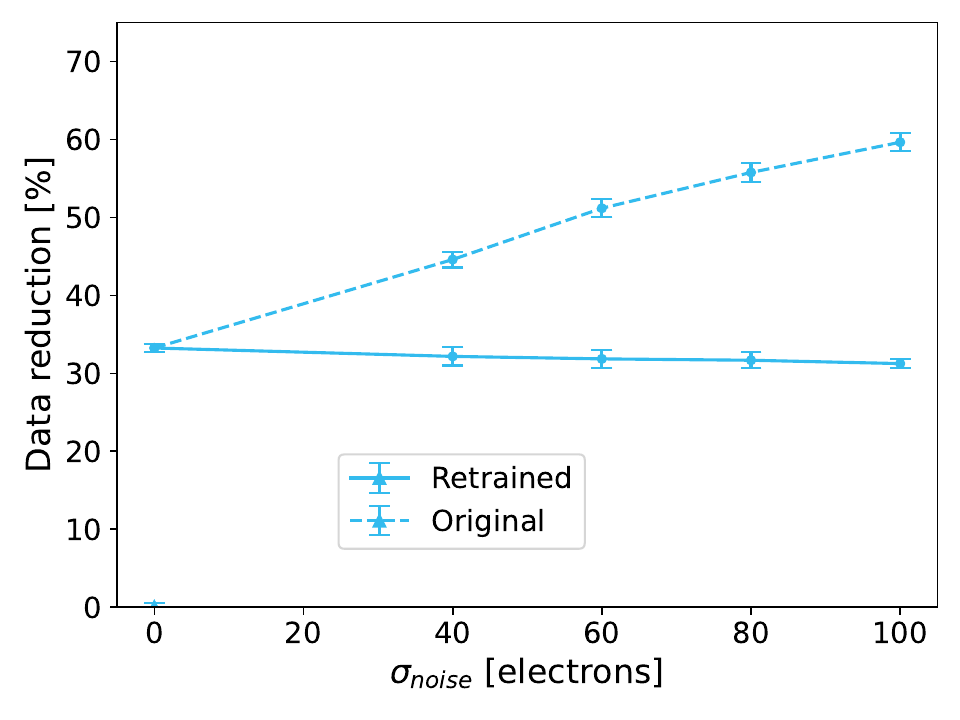}
    \caption{Performance results of the baseline 50$\times$12.5$\times$100~$\unit{\micro\meter}^3$ sensor geometry as a function of noise at $p_T$ boundary = 0.2~GeV. ``Original" corresponds to the model weights obtained by training on simulation with no Gaussian noise applied.}
    \label{fig:noise-study}
\end{figure}

The performance of the $p_T$ filtering algorithm is evaluated at four different noise levels up to 100 electrons, as shown in Figure \ref{fig:noise-study}. For each data point, noise sampled from a Gaussian distribution of $\mu$ = 0 and $\sigma = \sigma_{noise}$ was injected into the simulation datasets for further training/evaluation. We observe that there is a considerable drop in performance with increasing levels of noise. However, upon retraining the network on the noise-injected simulation datasets, the performance of the network can be retained. For instance, at $\sigma_\text{noise}$ = 100 electrons, the signal efficiency and data reduction of the retrained model lie within 3\% and 6\% to the original model's results at $\sigma_\text{noise}$ = 0 electrons, respectively.

To assess robustness to non-ideal detector conditions, we studied the impact of defective pixels (eventually to be masked), assuming a rate \textless~1\% \cite{pixel-masking-rate}, corresponding to approximately one pixel per 21$\times$13 array. A network was trained on datasets for the baseline sensor geometry, after injecting noise sampled from a Gaussian distribution (with $\mu$ = 0 electrons, $\sigma$ = 80 electrons) and imposing a per-pixel noise threshold of 400 electrons, with no defective or masked pixels. Its performance was then evaluated by introducing a masked (zeroed) pixel at different locations within the pixel array. The resulting performance remained within 1\% of the nominal case, indicating that the network is robust to the expected level of pixel masking during operation and does not require retraining.

We also evaluated the model performance on the radiation damaged sensor dataset (1100 fb$^{-1}$) with injected noise and a 400 electron minimum charge threshold per pixel for the baseline and CMS Phase-2 sensor geometries. The ENC values calculated from simulation for the baseline and CMS Phase-2 sensor geometries were 65 and 105 electrons, corresponding to $\sigma_\text{noise}$ values of 90 and 125 electrons, respectively. To assess model robustness under extreme operating conditions, we retrained the model on noise-injected, irradiated sensor datasets with a threshold. This retraining substantially preserved performance: for the baseline sensor geometry, the retrained model's signal efficiency and data reduction remained within 10\% and 9\% of the baseline model (trained on unirradiated, noiseless data). For the CMS Phase-2 geometry, these metrics were within 6\% and 4\%, respectively.

\section{Conclusions and outlook}

On-detector inference of incident particle characteristics and on-sensor data filtering for highly granular pixel detectors has been explored in previous work~\cite{Yoo_2024}, but only for a single benchmark sensor geometry. In this paper, we have systematically evaluated the performance of the baseline filtering network across variations in sensor geometry, Lorentz drift, radiation damage, and noise. The results presented in this paper pertain to prompt tracks, and future studies of the network’s performance on displaced tracks in future work will provide additional useful insight into the feasibility of such a technology for long-lived particles and other non-prompt sources. Our results highlight that smaller sensor pitch and increased thickness improve classification performance due to better resolved cluster features. Across the geometries studied, the signal efficiency ranges from 88\% to 95\%, and the data reduction ranges from 27\% to 33\%. The presence of Lorentz drift enhances the discriminating power of the classification algorithm. The network performance is worse for sensors in the endcap than for those in the barrel, and while tuning the p$_T$ boundary improves signal efficiency, it comes at the cost of an absolute $\mathcal{O}$(10\%) decrease in data reduction. Additionally, we observed that noise and radiation damage severely degrade filtering performance, but they can be mostly compensated through retraining. As the detector ages, periodic retraining will be required, highlighting the importance of having reprogrammable logic in the ASICs.

These findings inform design considerations for future detectors aiming to implement intelligent, on-sensor data reduction. The proof-of-concept from prior work, augmented by the results presented in this paper, also provides valuable direction for future R\&D under the \smartpixels project. For use in the first layer of a trigger system, at least two layers of smartpixels are needed. This paper has focused on the innermost layer, which has the most stringent bandwidth requirements and is most critical for $b$-jet tagging and impact parameter measurement. The performance of additional layers will be studied in future work, including the per-layer performance but also the track parameter resolution for \smartpixels as a part of a multi-layer tracking system. The expected tracking performance is critical to estimating the improvement in physics capabilities of a collider experiment integrating smartpixels. We are also extending simulations to account for scenarios one can expect to occur in the experiment, studying a variety of machine learning algorithms, and prototyping and characterizing new ASICs developed for \smartpixels. Detailed studies of the analog front-end and on-chip neural network performance can be found in~\cite{asic-testing-paper}. An important next step, building on these efforts, will be hybridizing a \smartpixels ASIC with a pixel sensor and demonstrating the performance in test beam.  Continued efforts have the potential to reduce data throughput while maintaining high physics performance, which is essential for the success of next-generation collider experiments.

\section*{Acknowledgements}
This work was completed using computing resources at the University of Chicago's Research Computing Center and the Fermilab Elastic Analysis Facility. We thank Burt Holzman (Fermilab) for computing support. C M and M A W are supported by NSF award PHY-2208803. C M, D S, and A T are supported by DOE Office of Science award DE-SC0023715 from Funding Opportunity Announcement for Artificial Intelligence Research for High Energy Physics, DE-FOA-0002705. M L and A R D are supported by the A3D3 HDR Institute through the NSF award PHY-2117997. M S N and D J are supported through the NSF cooperative agreement OAC-2117997 and the DOE Office of Science award DE-SC0023365. M S and P M are supported by NSF award PHY-2310072. K F D and E H are supported by the NSF CAREER Program through award 2443370, and K F D is additionally supported by the Neubauer Family Assistant Professor Program. E Y is supported by the University of Chicago's Quad Undergraduate Research Scholar program, and D A is supported by the University of Chicago's Sachs Fellowship. A B is supported by the Schmidt Sciences Foundation. J D and B W are supported by DOE Office of Science award DE-SC0026236. A C is supported by NSF award PHY-2150125. L G, A G, and G D G are supported by DOE Office of Science award DE-SCL0000038. K U, J P, and R S are supported by DOE Office of Science award DE-SC0010005. 

\bibliographystyle{elsarticle-num} 
\bibliography{references}{}
\end{document}